\documentclass[12pt]{iopart}

\usepackage{amsfonts}
\usepackage{amssymb}
\usepackage{graphicx}
\usepackage{epstopdf}
\usepackage{hyperref}
\usepackage{xcolor}

\newcommand{\ket}[1]{$| #1 \rangle$} 

\begin{document}

\title{Two-qubit logical operations in three quantum dots system}

\author{Jakub {\L}uczak, Bogdan R. Bu{\l}ka}

\address{Institute of Molecular Physics, Polish Academy of Sciences, ul. M. Smoluchowskiego
17, 60-179 Pozna{\'n}, Poland}
\ead{jakub.luczak@ifmpan.poznan.pl}
\vspace{10pt}

\begin{abstract}
We consider a model of two interacting always-on, exchange-only qubits for which
controlled phase ($CPHASE$), controlled NOT ($CNOT$),
quantum Fourier transform ($QFT$) and $SWAP$ operations can be implemented only in a
few electrical pulses in a nanosecond time scale. Each qubit is built of
three quantum dots (TQD) in a triangular geometry with three electron spins which are
always kept coupled by exchange interactions only.
The qubit states are encoded in a doublet subspace and are
fully electrically controlled by a voltage applied to gate
electrodes.
The two qubit quantum gates are realized by short electrical pulses which change the
triangular symmetry of TQD and switch on exchange interaction between the qubits. We
found an optimal configuration to implement the $CPHASE$ gate by a single pulse of the
order 2.3 ns. Using this gate, in combination with single qubit operations, we searched
for optimal conditions to perform the other gates: $CNOT$,  $QFT$ and $SWAP$.
Our studies take into account environment effects and leakage processes as well. The
results suggest that the system can be implemented for fault tolerant quantum
computations.
\end{abstract}

%
\vspace{2pc}

\section{Introduction}

The basic unit in  quantum computers is
a qubit, which can be physically realized in superconducting circuits \cite{dicarlo},
trapped ions \cite{tan}, photons \cite{kok, wei},
molecular magnets \cite{wedge}, or a single defect in diamonds \cite{dolde}. Recent
progress in experimental fabrication of semiconducting quantum dots
(QDs) makes them one of the most perspective for quantum computation \cite{zwanenburg}.
By standard lithographic methods one can achieve large arrays of
QDs which can work as multi qubit quantum register required in universal quantum
computations. Full control of the device can be performed purely
electrically by gate voltages applied to external electrodes. Sensitive methods of
detecting single electron dynamics in QD, for example by measuring
currents in quantum point contacts (QPC) \cite{elzerman} or in single electron
transistors (SET) \cite{lu} give the opportunity to read-out the qubit
states with very high accuracy.

Loss and DiVincenzo \cite{loss} proposed a spin qubit encoded in a single electron spin
in a QD which is characterized by longer coherence time than
a charge qubit. To encode the spin qubit one needs to apply an external magnetic field
which removes the spin degeneracy. The control of the qubit
states is performed  by the electron spin resonance (ESR) \cite{nowack}. Recently
the two-qubit logic gate was performed by Veldhorst {\it et al.} \cite{veldhorst} in
isotopically enriched silicon double quantum dot systems.
Another proposition is to encode the qubit in two spin states where a singlet state and
one of triplet states (S-T) correspond to the north and the
south pole of the
Bloch
sphere.  The qubit rotation around one of the axis on the Bloch sphere can be performed
in a nanosecond scale via a pure electrical control of the
exchange
interaction \cite{petta}. The rotation around the second axis can be induced by a
controlled dynamic nuclear polarization \cite{foletti} or by a
magnetic field difference between two sides of the double dot with integrated
micromagnets
\cite{wu}. There are propositions \cite{mehl,srinivasa2015} to build a
system of two interacting S-T qubits. Mehl {\it et al}. \cite{mehl} proposed the
high-fidelity entangling quantum gate in two S-T qubits mediated by one
quantum state from a quantum dot between them. The S-T qubits can be also coupled via an
exchange  \cite{klinovaja} and a capacitive interaction
\cite{srinivasa2015}.

One of the most promising concept is an exchange-only qubit encoded in a doublet subspace
of three spins \cite{divincenzo}. The advantage of this
proposal is easy control of the qubit states by purely electrical manipulations of the
exchange interactions between the spins. Moreover, the doublet
subspace is protected
from
decoherence processes \cite{dfs}. The exchange-only qubits can be encoded in three
quantum dots (TQD) with a linear configuration \cite{aers,laird10,gaudreau}, a triangular
arrangement \cite{hawrylak,srinivasa,bulka,luczak,luczak2016} or in
a double dot system with many levels \cite{shi}.
Initialization and one-qubit operations are
performed by electrical pulses applied to  gate electrodes and was already demonstrated
experimentally for the linear TQD (l-TQD) \cite{gaudreau} and
theoretically for the triangular TQD (t-TQD) \cite{luczak,luczak2016} (for a recent
review see \cite{russ}). Manipulation of the qubit
can be also done by an rf voltage applied to one of
gate electrodes in a resonant exchange qubit \cite{medford,taylor2013}. If the rf
excitation energy matches to the energy difference between the qubit states one can
observed the Rabi nutation on the Bloch sphere. The read-out of the qubit states can be
done by measurement of the current flowing through the system in the doublet
\cite{luczak} or
quadruplet \cite{amaha} blockade regime. Recently an always-on exchange qubit (AEON)
\cite{shim} was  presented in the linear TQD system. In such configuration all exchange
couplings are always kept on during the qubit operations, which differs from previous
concepts. An advantage of this proposal is performing the quantum logical operations at a
sweet spot in detuning parameters, where charge fluctuations are minimized
\cite{russ,medford,stopa}.

A very important
challenge is implementation of a two-qubit logical operations for which one of the most
effective is the controlled phase gate ($CPHASE$). This gate in
combination with single-qubit gates can be used as a circuit for any universal quantum
computation. Recently Doherty and Wardrop \cite{doherty,wardrop} showed theoretically how
to implement the $CPHASE$ gate by a single exchange pulse
in the resonant exchange qubit encoded in the linear TQD system. They estimated the gate
operation time at 21 ns. Pal {\it et al.} \cite{pal} proposed capacitively-coupled two
exchange-only qubits for which the CNOT gate can be performed by varying the level
splitting of individual qubits and the inter-qubit coupling time.

In this paper we consider two interacting AEON qubits each encoded in the triangular
configuration of TQD.
Earlier it was shown that the qubit
states are sensitive to breaking of the triangular symmetry \cite{hawrylak,bulka,luczak}.
Moreover, in the triangular TQD any qubit state on the Bloch sphere
can be easily generated by an adiabatic Landau--Zener transition \cite{luczak2016}, which
is in
contrast to the linear geometry where one of the poles of the Bloch sphere is favorable.
This gives opportunity to construct multi-qubit register where each qubit can be encoded
in a desired state. Recently Noiri {\it et al.} \cite{noiri} showed an experimental realization of the TQD system in the triangular geometry formed at a GaAs/AlGaAs hetero-interface. Applying a potential gate voltage between the dots they were able to control the tunnelling barriers. This experiment suggests that the symmetry of TQD as well as exchange interactions can be fully electrical controlled by tunable inter-dot tunnel couplings.

Our main purpose is to study the two-qubit operations as $CPHASE$, $CNOT$, quantum
Fourier transform ($QFT$) and $SWAP$ which can been done in few impulses only. First we will consider implementation of the $CPHASE$ gate
in two coupled triangular TQD systems and show that it can be performed by a single
electrical pulse only. Next this gate, in combination with the one
qubit operations, will be used to search a most optimal configuration to perform the
$CNOT$
and $QFT$ gates. In the previous paper \cite{luczak} we showed that one-qubit gates can
be
performed in a single step by a quick change of the symmetry of the triangular TQD
system.
This
gives the opportunity to implement the $CNOT$ in 3 pulses only. In
the
earlier paper, by DiVincenzo et al. \cite{divincenzo}, $CNOT$
required 19 pulses and by Shi et al. \cite{shi} -- 14 pulses. The $SWAP$ operation is
usually implemented by three $CNOT$ gates \cite{nielsen}. Moreover we will show how to
directly perform $SWAP$ by only two pulses: switching on the exchange interaction between
the qubits and simultaneously performing the Pauli X-gate.

The research of two-qubit logical operations is supplemented by an analysis of a role of
an environment which disturbs the quantum system and its control. In the exchange only
qubits, the main sources of the decoherence are the magnetic noise \cite{lidar2013} due
to nuclear spins and the charge noise \cite{russ2016} related with the random potential
fluctuations of the experimental set-up. For the single qubit those effects can be
suppressed by encoding the qubit in the DFS subspace \cite{dfs} and operating in the
sweet spot \cite{shim}. In this paper we focus on potential fluctuations breaking the
triangular symmetry of the system and their influence on two-qubit logical operations. We
estimate the fidelity and leakage of the two-qubit gates performed close to the optimal
conditions. These results are essential for implementation of the considered systems in
fault tolerant quantum computations.

The paper is organized as follows. Section 2 describes the model of single- and two-qubit system encoded in the triangular geometry of TQD. In Section 3 we study in details the two-qubit logical quantum operations. The analysis of the leakage and the fidelity of performed gates is presented in Sec. 4. Finally we conclude the paper in Section 5.

\section{Modeling of two-qubit system}

\begin{figure}
\centering
\includegraphics[angle=-90,width=0.7\textwidth]{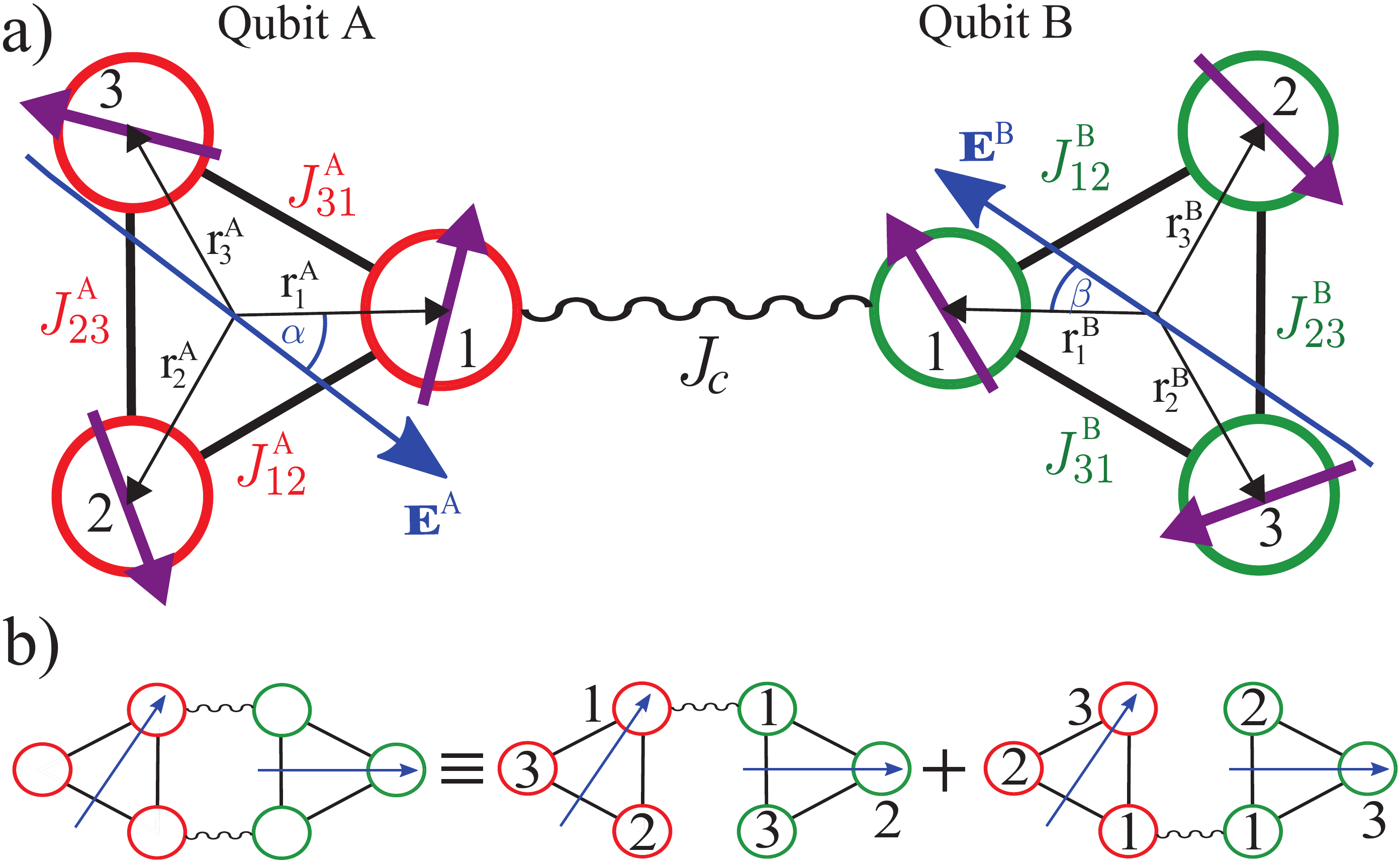}
\caption{a) Model of two interacting spin qubits built on two TQD systems in the presence
of the effective electric fields.  Panel b) presents the multi-connected TQD
system, which in the weak coupling limit can be modeled as a composition of two single
interactions. (Color figure online)}
\label{fig_model2TQD}
\end{figure}

We will consider two interacting exchange only spin qubits, each built on three
coherently coupled
quantum dots (TQD) in the triangular geometry - see
figure
\ref{fig_model2TQD}.

\subsection{Single qubit}

First, we briefly describe the single TQD system, which dynamics is governed by an
extended Hubbard Hamiltonian \cite{russ}
\begin{eqnarray}\label{hubbard}
  \hat{H} &=& \sum\limits_{i,\sigma}\epsilon_{i}\,n_{i\sigma}
+\sum\limits_{i,\sigma}t_{i,i+1}\left(c^{\dagger}_{i\sigma}c_{i+1\sigma}
  +h.c.\right)+\sum\limits_{i}U_in_{i\uparrow}n_{i\downarrow} \\ \nonumber
  &&+ \sum_i J^{dir}_{i,i+1}\left(\mathbf{S}_i\cdot\mathbf{S}_{i+1}- \frac{1}{4}\right)-g
\mu_B B_z \sum_i S_{z,i}\,,
\end{eqnarray}
where $\epsilon_{i}$ is a local site energy, $t_{i,i+1}$ is a hopping parameter between
the dots and $U_i$ describes a intra-dot Coulomb interaction. The direct interaction
$J^{dir}_{i,i+1}$ originates from a quantum exchange term of the Coulomb interaction
between electrons on the dots $i$ and $i+1$ \cite{yosida}. For a
defined confinement potential it can be calculated by means of the Heitler--London and
Hund--Mulliken method as a function of the interdot distance, the potential barrier and
the magnetic field \cite{burkard,li}. In experiments on exchange qubits these
parameters can be purely electrical controlled by potential voltages applied to the
quantum dots \cite{laird10,gaudreau}. It has been shown \cite{mizel} that for several
spins engaged in mutual interactions, both the quantitative and qualitative effects arise
which modify the standard form of the Heisenberg exchange interaction. The last term in
(\ref{hubbard}) corresponds to the Zeeman splitting by an external magnetic field $B_z$
($\mu_B$ is the Bohr magneton, \emph{g} is the electron g-factor).

We assume that the qubit
system is in the charge region (1,1,1), with one electron on each dot. Deep in this region there is the sweet spot, which can be easily achieved
experimentally be
proper shifting the local site energies $\epsilon_{i}$ \cite{aers, laird10, gaudreau}.
For the large Coulomb interaction $U\gg |t_{i,i+1}|$  one can use the canonical
transformation \cite{kostyrko}, which excludes the local two-electron states, to get an
effective spin Hamiltonian:
\begin{eqnarray}\label{heisenberg}
H= \sum_i J_{i,i+1}\left(\mathbf{S}_i\cdot\mathbf{S}_{i+1}- \frac{1}{4}\right)-g \mu_B
B_z \sum_i S_{z,i}\,.
\end{eqnarray}
Here, a total exchange interaction $J_{i,i+1}=J_{i,i+1}^{dir}+J_{i,i+1}^{kin}$ contains
the direct exchange, $J_{i,i+1}^{dir}$, and the Anderson kinetic exchange
$J_{i,i+1}^{kin}$ , which is derived within the second order perturbation theory as
$J^{kin}_{i,i+1}=4t_{i,i+1}^2U/[U^2-(\epsilon_i-\epsilon_{i+1})^2]$.

The Hamiltonian (\ref{heisenberg}) describes the AEON qubit for which the exchange
interactions are controlled electrically by the gate potentials $V_{i,i+1}$ applied between two neighbour dots (similarly as was done in a gate-defined TQD device in a GaAs 2DEG with tunable inter-dot tunnel barriers \cite{noiri}).
For the linear approximation  $J_{i,i+1}=J+j_V V_{i,i+1}$, where $j_V$ is a sensitivity
of the exchange coupling to the gate voltage $V_{i,i+1}$. This kind of the exchange
control can keep the qubit always in the sweet spot which  causes less charge fluctuation
than e.g. the detuning the local energy levels on each quantum dot \cite{shim}.

In this paper we are interested in the analysis of the symmetry breaking effects,
therefore it is more suitable to express the
gate voltages as an effective electric field ${\bf E}$.
For a small value of ${\bf E}$ the exchange couplings can be expressed as
\begin{eqnarray}\label{exchangeinE}
J_{i,i+1}=J+g_E\;\cos\left[\alpha+\left(i-\frac{1}{2}\right)\frac{2\pi}{3}\right],
\end{eqnarray}
where $g_E =e|{\bf E}||{\bf r}_1-{\bf r}_2|/2$, $\mathbf{r}_i$ is the
vector showing the position of the $i$-th quantum dot
and $\alpha$ is the angle between the vectors of electric field and $\mathbf{r}_1$.

For three spins in single TQD there are two possible subspaces, with the quadruplets and
the doublets. The quadruplet states with the total spin $S=3/2$
and
$S_z=\pm1/2,\pm3/2$ are given by:
\begin{eqnarray}\label{quadruplet}
|Q^{+1/2}\rangle=\frac{1}{\sqrt{3}}(|\uparrow_1\uparrow_2\downarrow_3\rangle
+|\uparrow_1\downarrow_2\uparrow_3\rangle+|\downarrow_1\uparrow_2\uparrow_3\rangle),
\\
|Q^{+3/2}\rangle=|\uparrow_1\uparrow_2\uparrow_3\rangle,
\end{eqnarray}
and similar functions for opposite spin orientations. Energy of these states is
$E_{Q}^{S_z}=- g \mu_B B_zS_z$.
The second subspace is formed by the doublet states with $S=1/2$ and $S_z=\pm1/2$. For
$S_z=+1/2$ we choose the basis:
\begin{eqnarray}\label{d0}
|0^{1/2}\rangle
&=&\frac{1}{\sqrt{2}}(|\uparrow_1\uparrow_2\downarrow_3\rangle-|\uparrow_1\downarrow_2\uparrow_3\rangle)
\equiv |\uparrow_1\rangle|S_{23}\rangle
\\
|1^{1/2}\rangle&=&\frac{1}{\sqrt{6}}(|\uparrow_1\uparrow_2\downarrow_3\rangle+|\uparrow_1\downarrow_2\uparrow_3\rangle
-2|\downarrow_1\uparrow_2\uparrow_3\rangle)
\nonumber \\
&\equiv&\frac{1}{\sqrt{3}}|\uparrow_1\rangle|T_{23}^0\rangle-\sqrt{\frac{2}{3}}|\downarrow_1\rangle
|T_{23}^{1} \rangle, \label{d1}
\end{eqnarray}
where
$|S_{ij}\rangle=(|\uparrow_i\downarrow_j\rangle-|\downarrow_i\uparrow_j\rangle)/\sqrt{2}$
is the singlet state and
$|T_{ij}^{0}\rangle=(|\uparrow_i\downarrow_j\rangle+|\downarrow_i\uparrow_j\rangle)/\sqrt{2}$,
$|T_{ij}^{+1}\rangle=|\uparrow_i\uparrow_j\rangle$ are
the triplet
states on the bond $ij$.
Similarly one can express the doublets for $S_z=-1/2$ reversing all spin orientations.

We assume that the qubit A (B) is encoded in the doublet subspace \ket{0^{1/2}} and
\ket{1^{1/2}}, Eqs. (\ref{d0})-(\ref{d1}). In further considerations the spin index is
omitted for simplification of the notation. The Heisenberg Hamiltonian (\ref{heisenberg})
in the qubit basis can be rewritten in the form, where the index A (B) have been added to
distinguish the qubits:
\begin{eqnarray}\label{matrix2}
{\cal{H}}^{A(B)}=-\frac{1}{2}(3J^{A(B)}+g \mu_B
B_z)\mathbf{I}+\frac{\delta^{A(B)}}{2}\sigma_z+\frac{\gamma^{A(B)}}{2}\sigma_x
\end{eqnarray}
where {\textbf I} is a identity matrix, $\sigma_x$ and $\sigma_z$ are the Pauli
matrices, and
\begin{eqnarray}\label{jjj}
J^{A} &=& \frac{1}{3}(J_{12}^{A}+J_{23}^{A}+J_{31}^{A}), \\ \label{delta}
\delta^{A} &=& \frac{1}{2}(J_{12}^{A}+J_{31}^{A}-2J_{23}^{A})\\ \nonumber &=&
\frac{3}{2}g_E^{A}\cos\alpha, \\ \label{gamma}
\gamma^{A} &=& \frac{\sqrt{3}}{2}(J_{12}^{A}-J_{31}^{A})=-\frac{3}{2}g_E^{A}\sin\alpha.
\end{eqnarray}
Notice that $\cal{H}^A$ describes the single qubit in an effective magnetic field
$\textbf{b}=(\gamma^{A}, 0,\delta^{A})$, with
$\gamma^{A}$ and $\delta^{A}$ to
be its the \emph{x} and \emph{z} component. The eigenvalues of $\cal{H}^A$ are:
\begin{eqnarray}
E^{A}_\pm=-\frac{3}{2}J^{A}-\frac{g \mu_B B^{A}_z}{2}\pm\frac{\Delta^{A}}{2},
\end{eqnarray}
where $\Delta^{A}=\sqrt{(\delta^{A})^2+(\gamma^{A})^2}=3 g_E^A/2$ is the doublet
splitting.  Taking parameters suitable for Si/SiGe quantum dots
\cite{shi2013} one can estimate $\Delta^{A} \approx 24.6\mu$eV. Similarly for qubit B.

The qubit initialization can be performed by an adiabatic Landau--Zener transition
\cite{laird10,luczak2016} to the charge region $(1,1,1)$ form a neighbour charge state.
In this passage one can control, by another set of potential gates, the exchange
couplings, and finally, reach the sweet spot with well defined the triangular symmetry of
the TQD system, with a given orientation of the electric field ${\bf E}$. If $\alpha=0$,
the electric field is oriented toward the dot 1, the qubit parameters are $\gamma^A=0$
and $\delta^A=-3/2g_E^A$, then the qubit is encoded in the state \ket{1} - pointing to
the south pole of the Bloch sphere. For the opposite orientation of the electric field,
$\alpha=\pi$, the parameter $\delta^A=3/2g_E>0$ which enables preparation of the qubit in
the state \ket{0} - pointing to the north pole. Notice, that this procedure always
provides initialization of the qubit in the ground state.

After the Landau--Zener passage to the sweet spot and the initialization of the qubit one
can perform one--qubit quantum gates - only a single step is needed which  changes the
symmetry of the system, i.e. a change of the angle $\alpha$ of the electric field ${\bf
E}$ \cite{luczak}. Taking $\alpha=\pi/2$ one can perform the rotation of the qubit state
around the {\it x} axis of the Bloch sphere. For this case the parameters $\delta^A=0$,
$\gamma^A=-3/2 g^A_E$ and from the solution of the Schr\"{o}dinger equation one can find
a unitary operator of the rotation around the {\it x} axis as: $U_x=\exp[-i \gamma^A
\sigma_x t/2]$. The Pauli X-gate can be performed in time $t_X=2\pi/\gamma^A$ for which
the qubit state is changed to the opposite one. Similarly, one can make the Pauli Z-gate
operation taking $\alpha=0$ for which the qubit state rotates around the {\it z} axis and
its evolution is governed by the operator $U_z=\exp[-i \delta^A \sigma_z t/2]$. If one
changes the angle $\alpha$ to $3\pi/4$ the parameters $\delta^A$ and $\gamma^A$ become
equal. It induces the rotation of the qubit state around the vector
$(-1/\sqrt{2},0,-1/\sqrt{2})$ which corresponds to the Hadamard gate with $U_H=\exp[-i
(\delta^A \sigma_z+\gamma^A \sigma_x) t/2]$ and the operation time $t_H=\pi/\Delta^A$.

The presented model is general and can describe the linear molecule as well. For the
linear molecule one of the exchange interactions $J^A_{i,i+1}$ is zero. In this case the
effective electric field cannot be oriented in the full angle, and the adiabatic
generation of the qubit is limited to one hemisphere of the Bloch sphere only. For
example for $J_{31}=0$, the electric field angle can take the values $0\leq \alpha\leq
4\pi/3$ and the encoded qubit is oriented to the north hemisphere (the parameter
$\delta^A<0$ and $\gamma^A\neq 0$). To encode the qubit on the south hemisphere one needs
to perform a diabatic Landau--Zener transition in which the excited state $E^A_+$ is
involved as well \cite{luczak2016}. One can also initialize the qubit in the ground state
and by proper sequences of pulses perform the Pauli X-gate \cite{hanson}. This method
requires at least three additional pulses which increases the operation time.

\subsection{Two interacting qubits}

Let us now consider two triangular TQD systems interacting with each other as presented
in Fig. \ref{fig_model2TQD}a. The total Hamiltonian can be expressed as
\begin{eqnarray}\label{hamil_full}
H^{tot}=H^{A}+H^{B}+H^{int}\;,
\end{eqnarray}
where the interaction term is confined to two neighbourhood spins in the system A and B
and is given by
\begin{eqnarray}\label{Hint}
H^{int}= J_{c}\left(\mathbf{S}^{A}_{1}\cdot\mathbf{S}^{B}_{1}- \frac{1}{4}\right)\,.
\label{H_int}
\end{eqnarray}
Here, $J_c$ is an exchange coupling parameter which can be controlled by a potential
gate applied to a tunnel barrier between the qubits similarly like controlling the inter-dot coupling \cite{noiri}.

The two-qubit basis is built from the doublet states (\ref{d0}) and (\ref{d1}), and can
be
written as:
\begin{eqnarray}
\{|00\rangle,|01\rangle,|10\rangle,|11\rangle\},
\label{q_basis}
\end{eqnarray}
where \ket{AB} corresponds to the state in the qubit A and B. The total spin of two qubit
state is $S=1$ with $S_z=+1$.
The total two qubit Hamiltonian can be written as
\begin{eqnarray}
{\cal{H}}^{tot}={\cal{H}}^{A}\otimes\mathbf{I}^{2\times2}+\mathbf{I}^{2\times2}\otimes{\cal{H}}^{B}
+{\cal{H}}^{int}\;,\label{H_tot}
\end{eqnarray}
where two first terms describe the single qubits A and B. For the electric fields
oriented toward the dots 1, i.e. for $\alpha=\beta=0$,  the
interaction part (\ref{Hint}) can be rewritten in the basis (\ref{q_basis}) as
\begin{eqnarray}\label{inter11}
{\cal{H}}^{int}(\alpha=0,\beta=0)=J_{c}\left[
\begin{array}{c c c c}
0 & 0 & 0 & 0 \\
0 & -1/3 & 0 & 0 \\
0 & 0 & -1/3 & 0 \\
0 & 0 & 0 & -2/9
\end{array}
 \right].
\end{eqnarray}
Using the rotation matrix technic \cite{matrixrot} this Hamiltonian can be generalized for any orientation of the electric fields
\begin{eqnarray}
{\cal{H}}^{int}(\alpha,\beta)=R^{-1}(\alpha,\beta){\cal{H}}^{int}(0,0) R(\alpha,\beta)\;,
\label{H_rot}
\end{eqnarray}
where $R(\alpha,\beta)=R(\alpha)\otimes R(\beta)$ is the rotation matrix for the two
qubit system. The rotation of the electric field in the single qubit A is given by
\begin{equation}
R(\alpha)=\left[
\begin{array}{cc}
  \cos(\alpha) & -\sin(\alpha) \\
  \sin(\alpha) & \cos(\alpha)
\end{array}
\right]
\end{equation}
and similarly for the qubit B.

The interaction Hamiltonian $H^{int}$ (\ref{Hint}) does not conserve the local spin numbers
in the qubits A and B, which causes leakage from the two-qubit operation space.  Therefore, we assume
that the inter-qubit coupling $J_c$ is small compared to the intra-qubit  couplings
$J_{i,i+1}^{A(B)}$ and two qubit dynamics can be well described by an interaction Hamiltonian ${\cal{H}}^{per}$ derived
in the lowest order perturbation theory with $J_c$ as a perturbative parameter. Latter, in Sec. 4, we will consider optimal conditions
for which leakage is below the threshold for quantum computing.

Using the Pauli matrix representation the perturbative
Hamiltonian  ${\cal{H}}^{per}$ can be expressed as
\begin{eqnarray}\label{per}\nonumber
{\cal{H}}^{per}(\alpha,\beta)&=&J_0\;\mathbf{I}^{4\times4}
+J_z(\sigma_z\otimes\mathbf{I}^{2\times2}+\mathbf{I}^{2\times2}\otimes\sigma_z) \\
&&+J_{zz}\sigma_z\otimes\sigma_z +
J_\bot\left(\sigma_x\otimes\sigma_x+\sigma_y\otimes\sigma_y \right).
\label{H_eff}
\end{eqnarray}
Here, all the parameters are functions of the angles $\alpha$ and $\beta$, and they are
proportional to $J_c$. $J_0$ and $J_z$ describe the shift of all two qubits levels and
the levels in the single qubits due to the perturbation by $J_c$. For two qubit logical
operations important parameters
are $J_{zz}$ and $J_\bot$ which describe the effective coupling between the qubits in the
$z$ direction and the $x-y$ plane.  Table~\ref{tab1} presents these parameters for some
specific angles $\alpha=(i-1)2\pi/3$ and $\beta=(j-1)2\pi/3$, when the electric fields
are
directed to the dot $i$ and $j$ in the qubit A and B, respectively. For example
${\cal{H}}_{1-1}^{per}$ corresponds to the Hamiltonian ${\cal{H}}^{per}(0,0)$
given by (\ref{inter11}).

\begin{table}
\caption{Two-qubit interaction parameters calculated in  the lowest order perturbation
approximation for identical qubits (with $\delta^A=\delta^B$,
$J^A=J^B$) expressed in the unit of $J_c$.
For different qubits ($|\delta^A-\delta^B|\gg J_c$) the parameter $J_{\bot}=0$ for
presented $i$-$j$, whereas other parameters are unaffected. The indexes $i$-$j$ denote
the
orientation of the electric fields towards to the dot $i$ and $j$ in the qubit A and B,
respectively.}
\begin{center}
\begin{tabular}{l|c|c|c|c}
 & ${\cal{H}}_{1-1}^{per}$& ${\cal{H}}_{2-2}^{per}$,
${\cal{H}}_{3-3}^{per}$&
 ${\cal{H}}_{1-2}^{per}$, ${\cal{H}}_{1-3}^{per}$&
${\cal{H}}_{2-3}^{per}$  \\
 \hline

$J_0$ & -2/9 & -2/9 & -13/72  & -2/9   \\

$J_z$ & 1/18 & -1/36 & 1/72  & -1/36   \\

$J_{zz}$ & 1/9 & 1/36 & -7/72  & 1/36   \\

$J_\bot$ & 0 & 1/24 & 0 & -1/24   \\
\end{tabular}
\label{tab1}
\end{center}
\end{table}

The $CPHASE$ gate demands   $J_{zz}\neq0$, moreover the
larger value results in a shorter operation time, therefore ${\cal{H}}_{1-1}^{per}$
is the optimal configuration for this gate. The case ${\cal{H}}_{2-2}^{per}$ has
non-zero $J_\bot$ and can be use to implement the $SWAP$ gate. The similar table, but for
the linear TQD configuration, can be found in \cite{doherty}.

The above analysis has been confined to the cases with a single connection between
the qubits, however, one can easily generalized it for a
multi-connected TQD system. When two TQD systems are connected by their bases then the
interaction Hamiltonian has similar form (\ref{per}) with the parameters being a sum of
those from Table~\ref{tab1} for an appropriate direction of the electric field $i$-$j$.
For example, for the
case presented in Fig. \ref{fig_model2TQD}b, the interacting Hamiltonian is
${\cal{H}}^{per}_{1-2}+{\cal{H}}^{per}_{3-3}$.  We expect that for
multi-connections the operation time should be shorter for some two-qubit gates.

\section{Two--qubit quantum gates}

Let us study the dynamics of two interacting qubits and show how to perform two-qubit
quantum logical operations like $CPHASE$, $CNOT$, $QFT$ and
$SWAP$. We would like to find most optimal schemes of control pulses which implement
these quantum gates. To this end we will consider the evolution of the
two-qubit state
\begin{eqnarray}
|\Psi(t)\rangle=a_{00}(t)|00\rangle+a_{01}(t)|01\rangle+a_{10}(t)|10\rangle
+a_{11}(t)|11\rangle
\end{eqnarray}
derived from the time dependent Schr\"{o}dinger equation for the Hamiltonian
${\cal{H}}^{tot}$ (\ref{H_tot}) with an appropriate series of pulses.

First, we focus on the $CPHASE(\varphi)$ gate, because it is one of the most universal
gates which in combination with one-qubit gates can be used to
perform any quantum algorithm. This gate is defined by the diagonal matrix \cite{benenti}
\begin{eqnarray}
CPHASE(\varphi)=diag\{1,1,1,\exp[\imath \varphi]\}\;,
\end{eqnarray}
where the phase $\varphi$ is added to the qubit B (the \emph{target qubit}) if
and only if the qubit A (the \emph{control qubit}) is in the state \ket{1}.
The gate can be performed by a single electrical pulse which switches on the exchange
interaction between two qubits. From Table \ref{tab1} one can see that the simpler
implementation of the $CPHASE(\varphi)$ gate can be done for the electric field
orientation $1$-$1$ or $1$-$2$ ($1$-$3$). In these cases the parameter $J_\bot=0$, and
therefore, the Hamiltonian (\ref{H_eff}) has an effective Ising form and the logical gate
is provided by the interaction $J_{zz}$. The operation time $t_{CP}$ can be estimated
from the condition $\varphi=\int_0^{t_{CP}}J_{zz} dt$ \cite{doherty,wardrop} and is
shortest for the largest $J_{zz}$. Table \ref{tab1} shows that the parameter $J_{zz}$ has
the largest value for
the orientation $1$-$1$, and thus it is the most optimal configuration to
perform the $CPHASE(\varphi)$ gate.
Taking $\varphi = \pi$, $J_c=0.1J^{A}$ and the exchange parameters $J^{A}=J^{B}=20\mu$eV
one can estimate the gate time as $t_{CP}= 9\pi/J_c \approx
2.3$ ns for Si quantum dots \cite{busl}.

\begin{figure}
\centering
\includegraphics[width=0.6\textwidth]{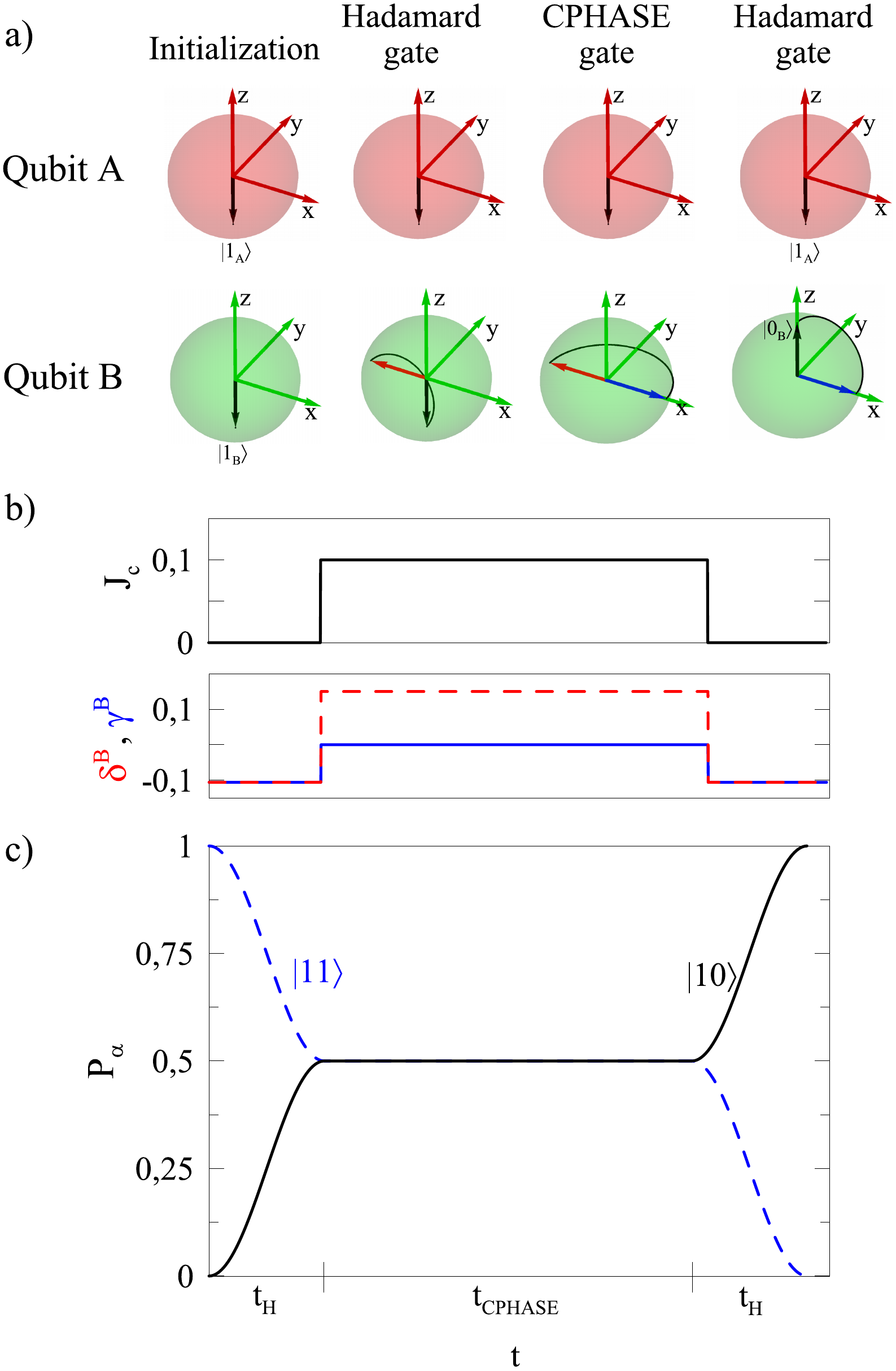}
\caption{Scheme of the realization of the $CNOT$ gate as a combination of the $CPHASE$
and
Hadamard gates performed from the initial state \ket{11} for the
$1$-$1$ orientation. Figure a)
presents a series of the operation steps on the Bloch spheres for the qubit B, while the
qubit A is unchanged. The middle panels b) show changes of $J_c$ and the parameters:
$\delta^B$ (red dashed line), $\gamma^B$ (blue solid line) after the sequence of voltage pulses.
Figure c) presents the probability of
occupation of the two-qubit states \ket{11} and \ket{10} as a function of time. $t_H$ and
$t_{CP}$ are the times needed to perform the Hadamard and the
$CPHASE$ gate, respectively. The exchange coupling $J_c=0.1$ and is switched on only
during
the $CPHASE$ operation, otherwise it is off. The other
parameters are $J^A=J^B=1$, $g_E^A=g_E^B=0.1$. (Color figure online)}
\label{cnot}
\end{figure}

$CPHASE(\varphi)$ for $\varphi=\pi$ in the combination with the Hadamard gates can be
used to perform the $CNOT$ gate. The $CNOT$ operation flips the
state on qubit B if and only if the qubit A is in state \ket{1} and is given by
\begin{eqnarray}
CNOT=H_{gate}^B~CPHASE(\pi)~H_{gate}^B\;,
\end{eqnarray}
where $H_{gate}^B$ is the Hadamard gate performed on the qubit B \cite{nielsen}. The
scheme of the realization of the $CNOT$ gate is presented in
Fig.~\ref{cnot}.
In the first step (see Fig.~\ref{cnot}a) the system is initialized in the state \ket{11}
(for the angle $\alpha=\beta=0$). Next, a single pulse is applied to the qubit B which
changes $\beta$ to $3\pi/4$ and the parameters $\delta^B$ and $\gamma^B$ become equal
(see fig~\ref{cnot}b). It
induces a rotation
around the vector $(-1/\sqrt{2},0,-1/\sqrt{2})$ on the Bloch sphere which corresponds to
the Hadamard gate.
Afterwards the interaction $J_c$ between the qubits is switched on and the $CPHASE(\pi)$
gate is performed in time $t_{CP}$, which is seen as the rotation around z-axis
over $\pi$ on the Bloch sphere. Finally one needs to apply another Hadamard gate to the
qubit B.
Notice that $J_c\neq0$ only during the $CPHASE(\pi)$ operation which reduces leakage
processes. Figure \ref{cnot}b presents the occupation probabilities
of the two-qubit states during the $CNOT$ operation. At the initial time the system is at
the state \ket{11} and the Hadamard gate transforms it to the superposition
$(|10\rangle-|11\rangle)/\sqrt{2}$.
The operation time  is $t_H=\pi/\Delta^B$, which can be estimated as $t_H\approx 0.1$ ns
taking $\Delta^B \approx 21.6\mu$eV for Si/SiGe quantum dots
\cite{shi2013}. Next the $CPHASE(\pi)$
operation transforms the system to the state $(|10\rangle+|11\rangle)/\sqrt{2}$ in
the time $t_{CP}= 9\pi/J_c \approx 2.3$ ns. Notice that the
whole $CNOT$ operation requires 3 pulses only.

We would like to consider the $SWAP$ operation which can be used e.g. in quantum
teleportation \cite{vaidman}. The gate swaps the qubit states
\ket{01}$\leftrightarrow$\ket{10} and can be implemented
by three $CNOT$ gates according to the scheme presented above.
However, for $J_{\bot}\neq0$ (e.g. for the orientation $2$-$2$) the $SWAP$ gate can be
performed directly in two pulses only. In the initialization step
the qubits are encoded in the state $|\Psi_{ini}\rangle=|01\rangle$ by taking
$\alpha=\pi$, $\beta=0$ for $g_E^A = g_E^B$. Next, the effective electric field ${\bf
E}^{B}$ is reversed by a single pulse in the qubit B ($\beta$ is changed to $\pi$), which
changes the qubit symmetry as well. For two decoupled qubits
($J_c=0$) this operation rotates the state of qubit B around $x$-axis on the Bloch sphere
which is equivalent to the Pauli $X$-gate \cite{luczak}.
However, if the interaction between the qubits is switched on by the second pulse
($J_c\neq0$) then the state on the qubit A simultaneously rotates around the
$x$-axis, which corresponds to the $SWAP$ operation. Notice that both pulses are applied
at
the same time. In this case the solution of the Schr\"{o}dinger
equation is
\begin{eqnarray}\label{swapO}
|\Psi(t)\rangle=e^{-\imath \phi_0 t}\cos(2 J_\bot t)|01\rangle -\imath e^{-\imath \phi_0
t}\sin(2 J_\bot t)|10\rangle,
\end{eqnarray}
where the phase factor $\phi_0=-3(J^A+J^B)/2-J_{zz}+J_0$.
One can easily find the
operation time $t_{SWAP}=\pi/(4 J_\bot)$, which for the considered
orientation $2$-$2$ is  $t_{SWAP}=6\pi/J_c$ and is estimated as $1.55$ ns for Si devices
\cite{busl}. The operation time can be even shorter for
multi-connected TQD systems. For example, for two triangles connected by their bases the
interaction Hamiltonian is the sum
${\cal{H}}_{2-2}^{per}+{\cal{H}}_{3-3}^{per}$, then the parameter
$J_{\bot}$ is twice larger which implies twice shorter operation time.

The scheme presented above can be implemented to the $SWAP$ family of gates, in general
for $SWAP^{1/m}$ with $m\geq 1$ \cite{fan, balakrishnan}. The most universal is
$\sqrt{SWAP}$, for $m=2$, because along with single qubit gates it can be used to any quantum
algorithm. Moreover, it was shown \cite{balakrishnan} that this gate is the perfect
entangler in the $SWAP^{1/m}$ family. From Eq. (\ref{swapO}) one can find the operation
time $t_{\sqrt{SWAP}}=\pi/(8 J_\bot)$, which is twice shorter than $t_{SWAP}$.

\begin{figure}
\centering
\includegraphics[width=0.6\textwidth]{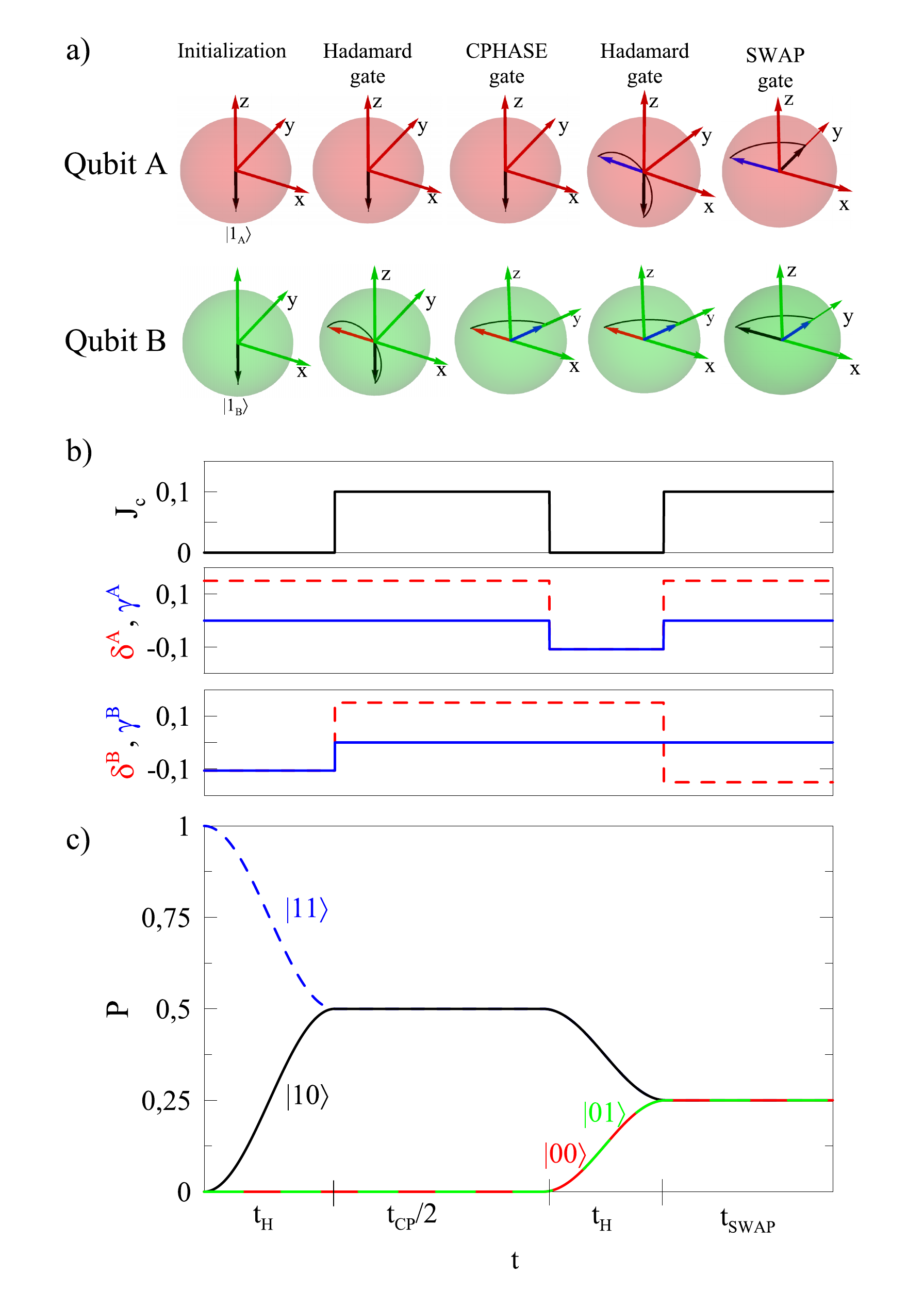}
\caption{Scheme of the realization of the $QFT$ operation as a combination of the
$CPHASE$, Hadamard gates and $SWAP$ performed from the initial state \ket{11}. Figure a)
presents a series of the operation steps on the Bloch spheres. Middle panels b)
show the time dependence of $J_c$, $\delta^{A(B)}$ and $\gamma^{A(B)}$ after the sequence of voltage pulses.
Figure c) presents the
probability of occupation of the two-qubit states as a function of
time. Notice, that the $CPHASE$ gate is performed in time $t_{CP}/2$. The parameters are the
same as in fig. \ref{cnot}. (Color figure online)}
\label{qft}
\end{figure}

As already mentioned, $CPHASE(\varphi)$ can be used in any quantum algorithms, e.g.
quantum factoring, quantum phase estimation for finding eigenvalues of a unitary operator
as well as the
order-finding problem. In all of these algorithm a key ingredient is the quantum Fourier
transform ($QFT$) \cite{nielsen}. $QFT$ is the unitary
operation for performing a Fourier transform of quantum mechanical amplitudes. The
quantum circuit
for the two-qubit $QFT$ can be expressed as \cite{nielsen}
\begin{eqnarray}
QFT=SWAP~H_{gate}^A~CPHASE(\pi/2)~H_{gate}^B
\end{eqnarray}
and its realization is presented in Fig. \ref{qft}. It is similar to the $CNOT$ gate,
however the first Hadamard gate
is performed on the qubit B whereas the second one on the qubit A. Moreover in the last step the $SWAP$ gate is applied.
Notice that the gate $CPHASE(\pi/2)$ changes the phase of the qubit B by
$\pi/2$, which means that the operation time is twice shorter. The total time needed for this gate can
be estimated as $t_{QFT}\approx 2.9$ ns.

\section{Leakage and Fidelity}

Let us now discuss in details the leakage which is a measure of how much of the initial qubit
state diffuses out of the logical qubit basis. For two interacting
TQD systems the total Hamiltonian $H^{tot}$, Eq.(\ref{hamil_full}), conserves the total
spin $S=1$ and its z-component $S_z=+1$.
However, the interaction Hamiltonian
$H^{int}$ does not preserve the local spin numbers in the qubits A and B, which leads to the leakage from the two-qubit space (\ref{q_basis}) during the logical operations.
There are 11 leakage states, which can be constructed from the
quadruplets \ket{Q^{S_z}} and the doublets \ket{0^{S_z}}, \ket{1^{S_z}} for the individuals qubits. From Table \ref{table2}, which presents the leakage states with corresponding energies, one can see that all two-qubit states are separated from the leakage states by the energy gap $\Delta E= 3J^{A(B)}/2$. Therefore, we expect that
the leakage is much suppressed in the considered triangular TQD system in comparison to the linear geometry
for with the two qubit state \ket{11} lies above four leakage states \cite{doherty, wardrop}.  For the weak coupling $J_c \ll \Delta E$ the leak is small, but on the other hand the gate operation time increases as $1/J_c$. Therefore, we need to find an optimal value of $J_c$ for fast gates and leakage errors below the threshold for fault tolerant computation.

\begin{figure}
\includegraphics[width=0.5\textwidth]{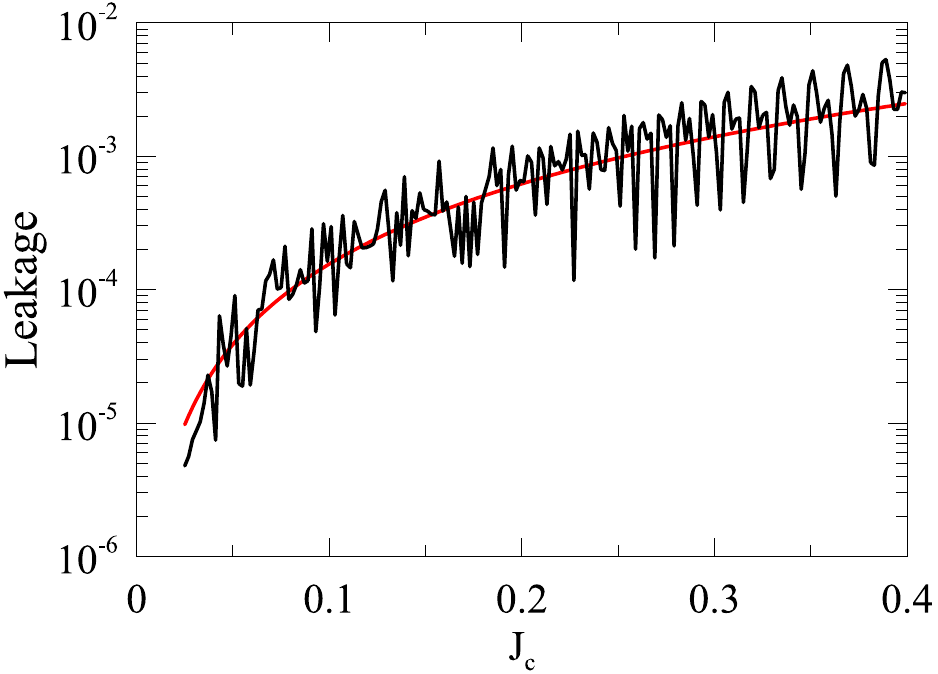}
\caption{The numerical results for the leakage as a function of inter-qubit coupling $J_c$. The red curve presents a dependence proportional to $(J_c/\Delta E)^2$. The calculations were performed for the orientation 1-1 between the qubits and after the $CPHASE(\pi)$ operation. The system parameters: $\alpha=\beta=0$ $g_E^A=0.1$, $g_E^B=0.2$, $J^A=J^B=1$. (Color figure online)}
\label{fig41}
\end{figure}

The leakage can be defined as \cite{wardrop, fazio}
\begin{eqnarray}
L=Tr\left(P\rho_r P \right),
\end{eqnarray}
where $P=1-\sum_{l=\{00,01,10,11\}}|l\rangle\langle l|$ is the projector off the computation
subspace and $\rho_r=|\Psi_r\rangle\langle\Psi_r|$ is the density matrix for the real final state \ket{\Psi_r} after the gate operation. Fig.~\ref{fig41} shows the numerical results of $L$ as a function of $J_c$ for the orientation 1-1 after the $CPHASE(\pi)$ operation. The level distribution in our case, Table 2, is similar to that one in Josephson qubits \cite{fazio}, for which the coupling to higher, leakage states can be regarded as a perturbation of the ideal two qubit system. The leakage in this limit can be estimated as proportional to $(J_c/\Delta E)^2$ \cite{fazio}, what is presented as a red curved in Fig.~\ref{fig41}. Since the threshold for
fault tolerant computation is estimated around $10^{-4}$ \cite{aliferis} our results suggest to take $J_c\approx0.1$ as the optimal value for two-qubit operations.

\begin{table}
\caption{Energy level diagram for two separate qubits in a magnetic field. The computation subspace is separated from the leakage subspace by the energy gap $\Delta E \approx 3J^{A(B)}/2$. }
\begin{center}
\begin{tabular}{l r}
\hline
leakage subspace & Energy \\
\ket{Q^{+1/2}Q^{+1/2}},\ket{Q^{+3/2}Q^{-1/2}},\ket{Q^{-1/2}Q^{+3/2}} &$0$ \\
\ket{1^{+1/2}Q^{+1/2}},\ket{1^{-1/2}Q^{+3/2}} & $-\frac{1}{2}(3J^A-\Delta^A)$ \\
\ket{Q^{+1/2}1^{+1/2}},\ket{Q^{+3/2}1^{-1/2}} & $-\frac{1}{2}(3J^B-\Delta^B)$ \\
\ket{0^{+1/2}Q^{+1/2}},\ket{0^{-1/2}Q^{+3/2}} & $-\frac{1}{2}(3J^A+\Delta^A)$ \\
\ket{Q^{+1/2}0^{+1/2}},\ket{Q^{+3/2}0^{-1/2}} & $-\frac{1}{2}(3J^B+\Delta^B)$ \\
\hline
2-qubit subspace & Energy \\
\ket{1^{+1/2}1^{+1/2}} & $-\frac{1}{2}[3(J^A+J^B)-\Delta^A-\Delta^B]$ \\
\ket{1^{+1/2}0^{+1/2}} & $-\frac{1}{2}[3(J^A+J^B)-\Delta^A+\Delta^B]$ \\
\ket{0^{+1/2}1^{+1/2}} & $-\frac{1}{2}[3(J^A+J^B)+\Delta^A-\Delta^B]$ \\
\ket{0^{+1/2}0^{+1/2}} & $-\frac{1}{2}[3(J^A+J^B)+\Delta^A+\Delta^B]$ \\
\hline
\end{tabular}
\label{table2}
\end{center}
\end{table}

The considerations above have been performed for the ideal generated qubits, with
precisely defined initial states, for the specific angels $\alpha$ and
$\beta$ of the electric fields ${\bf E}^{A}$ and ${\bf E}^{B}$ applied in the two-TQD
device. However in a real device there are many obstacles to reach these conditions, e.g.
an imperfect experimental arrangement or magnetic and charge noises. Since both qubits
are encoded in the doublet subspace which is the DFS the system is immune against global
magnetic field fluctuations. The other magnetic noise comes from the nuclear spins
surrounding the electron trapped in the quantum dot. This local magnetic field, called
the Overhauser field, causes leakage out of the computation states to subspace with $S_z\neq+1$. To suppress this
effect we assumed that both the Zeeman magnetic field and the exchange interaction are
larger than the Overhauser field. Because of the electrical control of the exchange-only
qubits the charge noise should be taken into consideration as well. It was shown
\cite{russ2016} that for a single qubit controlled by gating the potential barriers
between the quantum dots the sweet spot is located deep in the (1,1,1) regime, where the
estimated dephasing time is of the order of $\mu$s.

Our studies concern two AEON qubits, where small detuning of the exchange couplings
changes the triangular symmetry and all quantum operations are performed in the sweet
spot. Our aim is to consider mismatch of the angles $\alpha$ and $\beta$ of the electric
fields on the gate realization. If the angles are not precisely defined, e.g. caused by
fluctuations of the gate potentials, then the initial state can be in some superposition
with another qubit state. It has impact on realization of a logical operation and a final
two-qubit state.
The accuracy of the performed operation can be described by the fidelity which is a
measure
of the distance between two-quantum states, a desired ideal
state \ket{\Psi_{i}} and a real final state \ket{\Psi_{r}} after the gate operation
\cite{nielsen}. It can be expressed by
\begin{eqnarray}\label{fidelity}
F(\alpha,\beta)=Tr \left[\sqrt{\sqrt{\rho_{i}}\, \rho_{r}\, \sqrt{\rho_{i}}}\right]\,,
\end{eqnarray}
where $\rho_{i}=|\Psi_{i}\rangle\langle\Psi_{i}|$  and $\rho_r=|\Psi_r\rangle
\langle\Psi_r|$ are the corresponding density matrices. For the perfect
qubit gate the fidelity is unity, however in real devices it is lowered due an imperfect
setup of the initial state (the mismatch of $\alpha$ and
$\beta$) as well as leakage processes.

\begin{figure}
\includegraphics[width=0.9\textwidth]{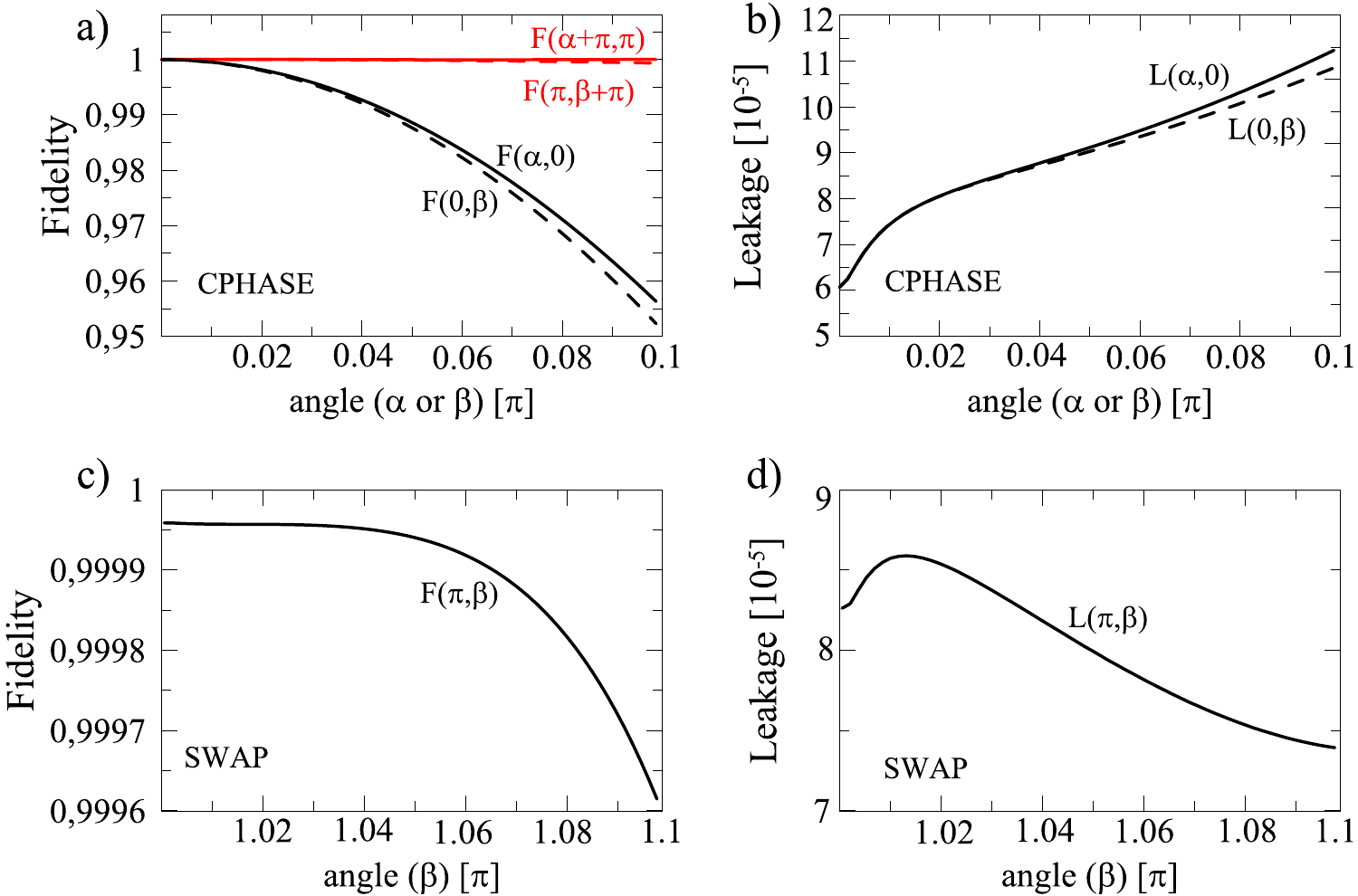}
\caption{Fidelity $F$ (a) and leakage $L$ (b) after the $CPHASE(\pi)$ gate plotted as a function of the angle $\alpha$ or $\beta$ (solid
or dashed curve) and  for two initial states \ket{11} and \ket{00}. The leakage $L=0$ for
to the state \ket{00}, and it is not presented. Bottom panel shows $F$ (c) and $L$ (d) after the SWAP gate as a function of $\beta$ for $\alpha=\pi$ for the initial state \ket{01}. In calculations we took $g_E^A=0.1$, $g_E^B=0.2$, $J^A=J^B=1$ and $J_c=0.1$. (Color figure online)}
\label{fig5}
\end{figure}

For the $CPHASE(\pi)$ gate the desired state is defined by
\begin{eqnarray}
|\Psi_{i}\rangle =
CPHASE(\pi)|\Psi_{ini}\rangle=\nonumber\\
a_{00}(0)|00\rangle+a_{01}(0)|01\rangle+a_{10}(0)|10\rangle
-a_{11}(0)|11\rangle.
\end{eqnarray}
 After the $CPHASE(\pi)$ operation all the coefficients should be unaffected,
$a_{ij}(t_{CP})=a_{ij}(0)$, except $a_{11}(t_{CP})=-a_{11}(0)$  which changes its
sign. Fig.~\ref{fig5}a) presents the fidelity for the $CPHASE(\pi)$ operation in the orientation 1-1
plotted as a function of the angle $\alpha$ (or $\beta$) for keeping $\beta$ (or
$\alpha$) fixed. At $\alpha=0$ the parameters
$\gamma^A = \gamma^B = 0$ and the initial state is $|\Psi_{ini}\rangle=|11\rangle$
precisely. In
this case the fidelity is $F(0,0)\approx 0.99994$. Its value is close to unity but is not
unity, because in the calculations we
included leakage processes as well.
If the initial state is not precisely generated due to a mismatch of the angle $\alpha$
($\beta$), then the parameter $\gamma^{A(B)} \neq 0$ and the single qubit is encoded in
the state \ket{1} with some contribution of the state \ket{0} which disturbs the
evolution of the qubit. One can observed a coherent rotation between the states
\ket{1} and \ket{0} with the oscillation period of the order
of $2\pi/\gamma^{A(B)}$ (comparable with $t_{CP}$). It affects the real
final state \ket{\Psi_r} and leads to a decrease of the fidelity $F$.
When the control qubit A is in state \ket{0}, then the $CPHASE(\pi)$ gate does not change
the state of the target qubit B. The two qubit system is much less sensitive to the
symmetry breaking in both the qubits and the fidelity is close to unity, $0.9992
<F<0.9997$ (see the red curves in Fig.~\ref{fig5}a).

Fig.~\ref{fig5}b presents the plot of the leakage $L$ after the $CPHASE(\pi)$ operation. We found that the state \ket{11}
is coupled to 7 leakage states. The largest contribution is due to the state
\ket{Q^{+1/2}Q^{+1/2}} for which $\langle11|H^{int}|Q^{+1/2}Q^{+1/2}\rangle=2/9J_c$  is the same order as those ones
in Table \ref{tab1}. However, the energy gap is large compared to $J_c$ (see Table \ref{table2}) which leads to the leakage $L\approx 6\cdot10^{-5}$,
see Fig. \ref{fig5}b). For the initial state \ket{00} the leakage $L=0$, because this state is not coupled to the leakage states.

We performed the fidelity and leakage calculations for the SWAP gate as well (see the bottom panel in Fig. \ref{fig5}). The results
show that this operation is less sensitive for mismatch of the angels $\alpha$
and $\beta$ and the fidelity $F\approx0.9998$. In this case
the leakage errors are in the order of $10^{-5}$, see fig. \ref{fig5}d, similarly like for the $CPHASE$ gate.
These results suggest that the considered TQD system can be implement
for fault tolerant computations even without applying noise corrections sequences
\cite{aliferis,hickman}.

\section{Conclusion}

The main result of this paper is presentation of a full set of single- and two-qubit quantum logical gates which can be easy performed in two AEON qubits each encoded in three coherently coupled quantum dots (TQD) in the triangular geometry. The implementation of the Pauli X-gate, the Pauli Z-gate, the Hadamard gate, $CPHASE$, $CNOT$, $QFT$ and $SWAP$ operations can be done in nanosecond time scale by few electrical pulses only. An advantage of the triangular geometry is simpler generation and encoding both of the qubit states by adiabatic Landau--Zener transitions which is in contrast to the linear AEON qubits \cite{shim}, where one state is preferable.

The single AEON qubit consists of three electrons and the qubit
states are encoded in the doublet subspace which is the decoherence-free
subspace \cite{dfs} and is immune against magnetic noises. Moreover each qubit operates always in the sweet spot, deep in the (1,1,1) charge region, where local charge noise on the quantum dots are suppressed. The triangular symmetry of the system
is fully controlled by short voltage pulses (in a scale of nanoseconds) which are applied
between the quantum dots to change exchange couplings. The one-qubit gates require only single pulses which is in contrast to linear AEON qubits where at least three pulses are needed \cite{shim}. For various orientations of the electric fields in both TQD systems we have
shown the most optimal configuration for
performing the two-qubit operations.

The $CPHASE(\varphi)$ gate can be implemented easily by means of a single electrical
pulse
which switch-on the interaction between the qubits. The most
optimal configuration for realization this gate is the configuration of the electric
field oriented to the dot 1 in both the qubits (see Fig.1) and the estimated operation
time for Si quantum dots is $t_{CP}\approx 2.3$ ns at $\varphi=\pi$. The $CPHASE(\pi)$
gate, in combination with two Hadamard
gates, are used to performed $CNOT$ gate. This gate requires 3
pulses only: the first and third one changes the symmetry of the second qubit and they
are used to perform the Hadamard gates, whereas the second pulse
performs the $CPHASE(\pi)$ gate. It is an advantage compared with the linear system
for which one needs 19 pulses to realize the $CNOT$ gate \cite{divincenzo}. A very
similar sequence of pulses can be used to perform the quantum Fourier transform, here
$CPHASE(\pi/2)$ -- changes the phase of the target qubit by $\pi/2$,  and the first
Hadamard gate
is applied to the qubit B whereas the second one to the qubit A.

We also showed how to perform the $SWAP$ gate. This operation requires the non-zero
coupling between the
qubits in the $x$-$y$ plane (with the parameter
$J_{\bot}\neq 0$) and the optimal initial configuration is for the orientation $2$-$2$.
This operation needs only two pulses and time $t_{SWAP}\approx 1.5$ ns.

Moreover, we considered the fidelity of the two-qubit operations which for $CPHASE(\pi)$ is very large $F\approx0.99994$ and $F\approx0.9997$ calculated at the
initial state \ket{11} and \ket{00}, respectively. When the control of the symmetry of
the system is disturbed by environment effects the fidelity is reduced but still very
high. We estimated the leakage during the gate operation  which is very small, the order
of $L \sim 10^{-5}$.

The triangular TQD systems with electrically tunable inter-dot tunnel couplings and local energy levels were already constructed \cite{noiri, rogge, seo}. We hope that the further technological progress in fabrication of semiconducting quantum dots enables in the near future to produce spin-qubit quantum registers based on the triangular TQDs and one can verify our theoretical predictions.

\ack This work has been supported by the National Science Centre, Poland, under the
project No. 2016/21/B/ST3/02160.


\section*{References}

\begin{thebibliography}{99}

\bibitem{dicarlo} DiCarlo L, Reed M D, Sun L, Johnson B R, Chow J M, Gambetta J M,
    Frunzio L, Girvin S M, Devoret M H,
    Schoelkop R J 2010 {\it Nature  Lett.} {\bf 467} 574.

\bibitem{tan} Tan T R, Gaebler J P, Lin Y, Wan Y, Bowler R, Leibfried D, Wineland D J
    2007 {\it Nature} {\bf 528} 380.

\bibitem{kok} Kok P, Munro W J, Nemoto K, Ralph T C, Dowling J P, Milburn G J 2007 {\it
    Rev. Mod. Phys.} {\bf 79} 135.

\bibitem{wei} Wei H-R, Deng F-G 2013 {\it Phys. Rev. A} {\bf 87} 022305; Wei H-R, Deng F-G, Long G L 2016 {\it Opt. Express} {\bf 24} 18619.

\bibitem{wedge} Wedge C J, Timco G A, Spielberg E T, George R E, Tuna F, Rigby S, McInnes
    E J L, Winpenny R E P, Blundell S J,
    Ardavan A 2012 {\it Phys. Rev. Lett.} {\bf 108} 107204.

\bibitem{dolde} Dolde F, Jakobi I, Naydenov B, Zhao N, Pezzagna S, Trautmann C, Meijer J,
    Neumann P, Jelezko F, Wrachtrup J 2013 {\it Nature Phys.} {\bf 9} 139.

\bibitem{zwanenburg} Zwanenburg F A, Dzurak A S, Morello A, Simmons M Y, Hollenberg L D C
    L, Klimeck G, Rogge S, Coppersmith S N, Eriksson M A 2013 {\it Rev. Mod. Phys.} {\bf
    85} 961.

\bibitem{elzerman} Elzerman J M, Hanson R, Willems van Beveren L H, Witkamp B,
    Vandersypen L M K, Kouwenhoven L P 2004 {\it Nature} {\bf 430} 431.

\bibitem{lu} Lu W, Ji Z, Pfeiffer L, West K W, Rimberg A J 2003 {\it Nature} {\bf 423}
    422.

\bibitem{loss} Loss D, DiVincenzo D P 1998 {\it Phys. Rev. A} {\bf 57} 120.

\bibitem{nowack} Nowack K C, Koppens F H L, Nazarov Yu V, Vandersypen L M K 2007 {\it
    Science} {\bf 318} 1430.

\bibitem{veldhorst} Veldhorst M, Yang C H, Hwang J C C, Huang W, Dehollain J P, Muhonen J
    T, Simmons S, Laucht A, Hudson F E, Itoh K M, Morello A, Dzurak A S 2015 {\it Nature}
    {\bf 526} 410.

\bibitem{petta} Petta  J R, Johnson  A C, Taylor J M, Laird E A, Yacoby A, Lukin M D,
    Marcus C M, Hanson M P, Gossard A C 2005 {\it Science} {\bf 309} 2180.

\bibitem{foletti} Foletti S, Bluhm H, Mahalu D, Umansky V, Yacob A 2009 {\it Nature
    Phys.} {\bf 5} 903.

\bibitem{wu} Wu X, Ward D R, Prance J R, Kim D, Gamble J K, Mohr T R, Shi Z, Savage D E,
    Lagally M G, Friesen M, Coppersmith S N, Eriksson M A 2014 {\it Proc. Natl. Acad.
    Sci. USA} {\bf 111} 11938.

\bibitem{mehl} Mehl S, Bluhm H, DiVincenzo D P 2014 {\it Phys. Rev. B} {\bf 90} 045404.

\bibitem{srinivasa2015} Srinivasa V, Taylor J M 2015 {\it Phys. Rev. B} {\bf 92} 235301.

\bibitem{klinovaja} Klinovaja J, Stepanenko D, Halperin B I, Loss D 2012 {\it Phys Rev.
    B} {\bf 86} 085423.

\bibitem{divincenzo} DiVincenzo D P, Bacon D, Kempe J, Burkard G, Whaley K B 2000 {\it
    Nature (London)} {\bf408} 339.

\bibitem{dfs} Lidar D A, Chuang I L, Whaley K B 1998 {\it Phys. Rev. Lett.} {\bf 81}
    2594;
Bacon D, Kempe J, Lidar D A, Whaley K B 2000 {\it Phys. Rev. Lett.} {\bf 85} 1758.

\bibitem{aers} Aers G C, Studenikin S A, Granger G, Kam A, Zawadzki P, Wasilewski Z R,
    Sachrajda A S 2012 {\it Phys. Rev. B} {\bf 86} 045316.

\bibitem{laird10} Laird E A, Taylor J M, DiVincenzo D P, Marcus C M, Hanson M P, Gossard
    A C 2010 {\it Phys. Rev. B} {\bf 82} 075403.

\bibitem{gaudreau} Gaudreau L, Granger G, Kam A, Aers G C, Studenikin S A, Zawadzki P,
    Pioro-Ladriere M, Wasilewski Z R, Sachrajda A S 2012 {\it Nature Phys.} {\bf 8} 54.

\bibitem{hawrylak} Hawrylak P, Korusinski M 2005 {\it Solid State Commun.} {\bf 136} 508.

\bibitem{srinivasa} Srinivasa V, Levy J 2009 {\it Phys. Rev. B} {\bf 80} 024414.

\bibitem{bulka} Bu{\l}ka B R, Kostyrko T, {\L}uczak J 2011 {\it Phys. Rev. B} {\bf83}
    035301.

\bibitem{luczak} {\L}uczak J, Bu{\l}ka B R 2014 Phys. Rev. B {\bf 90} 165427.

\bibitem{luczak2016} {\L}uczak J, Bu{\l}ka B R 2017 {\it Quantum Inf Process} {\bf 16}
    10.

\bibitem{shi} Shi Z, Simmons C B, Prance J R, Gamble J K, Koh T S, Shim Y-P, Hu X, Savage
    D E, Lagally M G, Eriksson M A, Friesen M, Coppersmith S N 2012 {\it Phys. Rev.
    Lett.} {\bf 108} 140503.

\bibitem{russ} Russ M, Burkard G 2017 {\it J. Phys.: Condens. Matter} {\bf 29} 393001.

\bibitem{medford} Medford J, Beil J, Taylor J M, Rashba E I, Lu H, Gossard A C, Marcus C
    M 2013 {\it Phys. Rev. Lett.} {\bf 111} 050501.

\bibitem{taylor2013} Taylor J M, Srinivasa V, Medford J 2013 {\it Phys. Rev. Lett.} {\bf
    111} 050502.

\bibitem{amaha} Amaha S, Izumida W, Hatano T, Teraoka S, Tarucha S, Gupta J A, Austing D
    G 2013 {\it Phys. Rev. Lett.} {\bf 110} 016803.

\bibitem{shim} Shim Y-P, Tahan C 2016 {\it Phys. Rev. B} {\bf 93} 121410(R).

\bibitem{stopa} Stopa M, Marcus C M 2008 {\it Nano Lett.} {\bf8} 1778.

\bibitem{doherty} Doherty A C, Wardrop M P 2013 {\it Phys. Rev. Lett.} {\bf 111} 050503.

\bibitem{wardrop} Wardrop M P, Doherty A C 2016 {Phys. Rev. B} {\bf 93} 075436.

\bibitem{pal} Pal A, Rashba E I, Halperin B I 2015 {\it Phys. Rev. B} {\bf 92} 125409.

\bibitem{noiri} Noiri A, Kawasaki K, Otsuka T, Nakajima T, Yoneda J, Amaha S, Delbecq M
    R, Takeda K, Allison G, Ludwig A, Wieck A D, Tarucha S 2017 {\it Semicond. Sci.
    Technol.} {\bf 32} 8.

\bibitem{nielsen} Nielsen M A, Chuang I L 2010 {\it Quantum Computation and Quantum
    Information} (Cambridge University Press, Cambridge).

\bibitem{lidar2013} Lidar D, Brun T 2013 {\it Quantum Error Correction} (Cambridge:
    Cambridge University Press).

\bibitem{russ2016} Russ M, Ginzel F, Burkard G 2016 {\it Phys. Rev. B} {\bf 94} 165411.

\bibitem{yosida} Yosida K 1996 {\it Theory of Magnetism} (Springer--Verlag Berlin
    Heidelberg)

\bibitem{burkard} Burkard G, Loss D, DiVincenzo D P 1999 {\it Phys. Rev. B} {\bf 59}
    2070.

\bibitem{li} Li Q, Cywinski {\L}, Culcer D, Hu X, Das Sarma S 2010 {\it Phys. Rev. B}
    {\bf 81} 085313.

\bibitem{mizel} Mizel A, Lidar D A 2004 {\it Phys. Rev. Lett.} {\bf 92} 077903; Mizel A,
    Lidar D A 2004 {\it Phys. Rev. B} {\bf 70} 115310.

\bibitem{kostyrko} Kostyrko T, Bu{\l}ka B R 2011 {\it Phys. Rev. B} {\bf 84} 035123.

\bibitem{shi2013} Shi Z, Simmons C B, Ward D R, Prance J R, Mohr R T, Koh T S, Gamble J
    K, Wu X, Savage D E, Lagally M G, Friesen M, Coppersmith S N, Eriksson M A 2013 {\it
    Phys. Rev. B} {\bf 88} 075416.

\bibitem{hanson} Hanson R, Burkard G 2007 {\it Phys. Rev. Lett.} {\bf 98} 050502.

\bibitem{matrixrot} Baker A 2003 {\it Matrix Groups: An Introduction to Lie Group Theory} (Springer International Publishing AG).

\bibitem{benenti} Benenti G, Casati G, Strini G 2005 {\it Principles of Quantum Computation and Information} (World Scientific Publishing).

\bibitem{busl} Busl M, Granger G, Gaudreau L, S\'{a}nchez R, Kam A, Pioro-Ladriere M,
    Studenikin S A, Zawadzki P, Wasilewski Z R, Sachrajda A S, Platero G 2013 {\it Nature
    Nanotech.} {\bf 8} 261.

\bibitem{vaidman} Vaidman L, Yoran N 1999 {\it Phys. Rev. A} {\bf 59} 116.

\bibitem{fan} Fan H, Roychowdhury V, Szkopek T 2005 {\it Phys. Rev. A} {\bf 72} 052323.

\bibitem{balakrishnan} Balakrishnan S, Sankaranarayanan R 2008, {\it Phys. Rev. A} {\bf
    78} 052305.

\bibitem{fazio} Fazio R, Palma G M, Siewert J 1999 {\it Phys. Rev. Lett.} {\bf 83} 5385.

\bibitem{aliferis} Aliferis P, Gottesman D, Preskill J 2008 {\it Quant. Inf. Comp.} {\bf
    8} 0181.

\bibitem{hickman} Hickman G T, Wang X, Kestner J P, Das Sarma S 2013 {\it Phys. Rev. B}
    {\bf 88} 161303.

\bibitem{rogge} Rogge M C, Haug R J 2008 {\it Phys. Rev. B} {\bf 77} 193306; Rogge M C, Haug R J 2008 {\it Phys. Rev. B} {\bf 78} 153310.

\bibitem{seo} Seo M, Choi H K, Lee S-Y, Kim N, Chung Y, Sim H-S, Umansky V, Mahalu D 2013 {\it Phys. Rev. Lett.} {\bf 110} 046803.



\end{thebibliography}
\end{document}